# Frozen states in shear-induced coagulation: When history matters

Mingliang Xie

State Key Laboratory of Coal Combustion, Huazhong University of Science and Technology, Wuhan 430074, China

Corresponding author's Email: mlxie@mail.hust.edu.cn

**Abstract**：We discover a frozen state in decaying turbulent coagulation where the moment ratio $M_C$ stabilizes at 4.5, defying the theoretical prediction of 2 for homogeneous systems. This persistent state emerges from historical memory effects that freeze spatial correlations, challenging the reduction of spatially extended systems to zero-dimensional models. Our findings fundamentally reshape the understanding of long-term coagulation dynamics with immediate implications for aerosol science and industrial processing.

**Significance Statement:** This work reveals a new fundamental behavior in particle-laden flows where systems become trapped in non-equilibrium states due to persistent spatial correlations. The discovery challenges classical coagulation theory and necessitates a paradigm shift in how we model turbulent coagulation processes across atmospheric science, industrial applications, and environmental engineering.

**Nomenclature**
$M_C = M_0 M_2 / M_1^2$, normalized moment ratio;
$M_k$, $k$-th moment of particle size distribution;
$Re$, Reynolds number;
$Sc$, Schmidt number;
$Da$, Damköhler number;
$v$, kinematic viscosity;
$D_v$, particle diffusion coefficient;
$\beta_0$, characteristic coagulation rate;
**u**, velocity field;
$G$, shear rate;
$n$, particle number density
$v$, particle volume
$t$, time
$x, y$, Cartesian coordinates

## 1. Introduction

Particle coagulation in turbulent flows governs numerous natural and industrial processes, from atmospheric aerosol dynamics to powder processing (**Friedlander, 2000**). While the population balance equation (PBE) provides the fundamental framework for describing particle size distribution evolution, its direct solution remains computationally prohibitive (**Liao and Lucas, 2010**). Moment methods offer an efficient alternative by tracking statistical moments, dramatically reducing computational cost while preserving essential physics (**Xie, 2013**).

A cornerstone of coagulation theory is the concept of self-similar solutions (**Friedlander, 1966**), where normalized size distributions become invariant under scaling transformations (**Xie, 2023**). For constant coagulation kernels, the self-similar distribution corresponds to $M_C = 2$ – a prediction extensively validated in homogeneous systems (**Schumann, 1940**). However, real-world coagulation typically occurs in non-uniform flows where spatial transport interacts with coagulation dynamics (**Xie, 2025, Pan et al., 2025; Xie and Yang, 2025**). The common practice of spatial averaging implicitly assumes equivalence to homogeneous systems, but this assumption remains largely unexplored for nonlinear coagulation processes.

Here, we demonstrate that this assumption fails dramatically, leading to unexpected frozen states where system becomes trapped far from the expected equilibrium. Focusing on shear-induced coagulation in a decaying Taylor-Green vortex, we employ the Taylor-expansion moment method (TEMOM) (**Yu et al., 2008**) and average kernel method (AK) (**Pan et al., 2024**) with high-order numerical schemes (**Lele, 1992**). Surprisingly, $M_C$ stabilizes at approximately 4.5 rather than the predicted 2, even as flow decays and transport diminishes. This frozen state represents new fundamental behavior in particle-laden flows with profound implications for modeling and prediction.

## 2. Mathematical Formulation

### 2.1 Dimensionless parameters and characteristic scales

We introduce the characteristic scales listed in **Table 1**. The key dimensionless parameters are: Reynolds number ($Re = UL/\nu$), Schmidt number ($Sc = \nu/D_v$) and Damköhler number ($Da = \beta_0 n_0 T$), where $\nu$ is the kinematic viscosity, $D_v$ is the particle diffusion coefficient (size-dependent via Stokes-Einstein relation), and $\beta_0$ is the characteristic coagulation rate. All equations are dimensionless unless specified.

Table 1. Characteristic scales

| Quantity | Characteristic Scale | Definition |
| --- | --- | --- |
| Length | $L$ | Domain size |
| Time | $T$ | Flow time scale |
| Velocity | $U$ | Characteristic flow velocity |
| Particle volume | $v_0$ | Initial mean particle volume |
| Number density | $n_0$ | Initial number density |

## 2.2 Flow configuration

We consider a two-dimensional Taylor-Green vortex flow over $[-\pi, \pi] \times [-\pi, \pi]$ with velocity components $\mathbf{u}(u, v)$:

$$u = +\sin(x)\cos(y) \cdot f(t) \tag{1a}$$
$$v = -\cos(x)\sin(y) \cdot f(t) \tag{1b}$$

where $f(t) = \exp(-2t/Re)$ represents the temporal decay. The shear rate is:

$$G = 2|cos(x)||cos(y)| \cdot f(t) \tag{1c}$$

## 2.3 Population balance equation

The particle size distribution $n(v, x, t)$ evolves according to:

$$\frac{\partial}{\partial t}n(v,x,t) + \nabla \cdot (\mathbf{u}n) = \frac{1}{ReSc}\nabla \cdot (D\nabla n) + Da\left[\frac{\partial n}{\partial t}\right]_{coag} \tag{2}$$

with coagulation term:

$$\left[\frac{\partial n}{\partial t}\right]_{coag} = \frac{1}{2}\int_0^v \beta(v-v', v')n(v-v')n(v')dv' - n(v)\int_0^\infty \beta(v, v')n(v')dv' \tag{3}$$

Defining the $k$-th moment as $M_k(x,t) = \int_0^\infty v^k n(v,x,t)dv$, we obtain:

$$\frac{\partial M_k}{\partial t} + \nabla \cdot (\mathbf{u}M_k) = \frac{1}{ReSc}\nabla \cdot (D\nabla M_k) + Da \cdot S_k, \quad (k = 0,1,2) \tag{4}$$

For shear-induced coagulation with kernel $\beta(v, v') = \frac{G}{\pi}(v^{1/3} + v'^{1/3})^3$, the source terms become:

$$S_0 = -\frac{1}{2}\bar{\beta}M_0^2, \quad S_1 = 0, \quad S_2 = \bar{\beta}M_1^2 \tag{5}$$

with averaged kernel $\bar{\beta} = pv_a^q$, where $p = 8G/\sqrt{\pi}$, $q = 1$, and $v_a = M_1/M_0$ (**Xie, 2024**).

## 2.4 Boundary and initial condition

Initial moment fields are spatially uniform: $M_0(x,y,0) = M_{00} = 1.0$, $M_1(x,y,0) = M_{10} = 1.0$, $M_2(x,y,0) = M_{20} = 4/3$, corresponding to $M_C(t=0) = 4/3$ for polydisperse distribution. Periodic boundary conditions apply to all transported quantities.

### Theorem 1 (Conservation of $M_1$)

For shear-induced coagulation in a periodic domain $\Omega$ with incompressible flow and uniform initial condition $M_1(x, y, 0) = M_{10}$, the first moment $M_1$ remains spatially uniform and temporally constant: $M_1(x, y, t) = M_{10}, \forall (x,y) \in \Omega, t \geq 0$.

### Proof:

Under the specified conditions: (1) $S_1 = 0$ (mass conservation), (2) Initial uniformity, (3) Incompressibility ($\nabla \cdot \mathbf{u} = 0$), and (4) Periodic boundaries, the constant solution $M_1(x,y,t) = M_{10}$ satisfies equation (4) uniquely.

## 2.5 Normalized moment ratio

We focus on the evolution of:

$$M_C = \frac{M_0 M_2}{M_1^2} \tag{6}$$

which characterizes distribution breadth. The time evolution decomposes as:

$$\frac{\partial M_C}{\partial t} = T_{conv} + T_{diff} + T_{coag} \tag{7}$$

with transport terms:

$$T_{conv} = -\frac{1}{M_1^2}\left[M_2 \nabla \cdot (\mathbf{u} M_0) + M_0 \nabla \cdot (\mathbf{u} M_2)\right] + 2\frac{M_0 M_2}{M_1^3} \nabla \cdot (\mathbf{u} M_1) \tag{8a}$$

$$T_{diff} = \frac{1}{ReSc}\left\{\frac{1}{M_1^2}\left[M_2 \nabla \cdot (D\nabla M_0) + M_0 \nabla \cdot (D\nabla M_2)\right] - 2\frac{M_0 M_2}{M_1^3} \nabla \cdot (D\nabla M_1)\right\} \tag{8b}$$

$$T_{coag} = Da\bar{\beta} M_0 \left(1 - \frac{1}{2} M_C\right) \tag{8c}$$

The effective diffusion coefficient in TEMOM is (**Xie, 2025**):

$$D = \frac{2M_C + 7}{9} \cdot v_a^{-\frac{1}{3}} \tag{9}$$

For homogeneous shear-induced coagulation, analytical solution yield:

$$M_0(t) = M_{00} \, \exp\left[\frac{G_0 Re M_{10}}{\sqrt{\pi}}\left(e^{-\frac{2t}{Re}} - 1\right)\right] \tag{10a}$$

$$M_1(t) = M_{10} \tag{10b}$$

$$M_2(t) = M_{20} - \frac{2M_{10}^2}{M_{00}} + \frac{2M_{10}^2}{M_0(t)} \tag{10c}$$

As $t \to \infty$, $M_0(t) \to 0$ and $M_C \to 2$, confirming the classical self-similar state.

### 2.6 Numerical methodology

We employ fourth-order Pade schemes for spatial discretization and fourth-order Runge-Kutta method for time integration (**Lele, 1992; Xie, 2025; Pan et al., 2025**). Comprehensive grid convergence studies determined $512 \times 512$ resolution (changes < 0.1% vs $1024 \times 1024$) with $\Delta t = 10^{-4}$ (CFL < 0.3, Fourier number < 0.25) ensuring balanced accuracy and efficiency.

## 3. Results and discussions

### *3.1 The frozen state phenomenon*

Our simulations reveal a striking departure from classical coagulation theory: rather than approaching $M_C = 2$, the system stabilizes at $M_C \approx 4.5$, forming a frozen state that persists indefinitely.

### *Theorem 2 (Vanishing transport contributions)*

For a periodic domain with sufficiently smooth moment fields, spatially averaged convective and diffusive transport contributions vanish exactly:

$$\langle T_{conv} \rangle = 0, \qquad \langle T_{diff} \rangle = 0$$

where $\langle \cdot \rangle$ denotes spatial averaging over the domain.

***Proof:***

Following divergence theorem and the periodic boundary conditions, all divergence terms $\langle \nabla \cdot (\mathbf{u} M_k) \rangle$ and $\langle \nabla \cdot (D \nabla M_k) \rangle$ vanish identically.

This theorem establishes that only coagulation drive evolution of the spatially averaged $M_C$. In. the asymptotic regime ($\langle T_{conv} \rangle \to 0$, $\langle T_{diff} \rangle \to 0$):

$$\frac{\partial \langle M_C \rangle}{\partial t} = \langle T_{coag} \rangle = \langle \bar{\beta} M_0 \rangle \left(1 - \frac{M_C}{2}\right) \approx 0 \tag{11}$$

Homogeneity would suggest $\langle M_C \rangle \to 2$, but persistent spatial correlations modify the dynamics:

$$\langle M_0 M_2 \rangle = \langle M_0 \rangle \langle M_2 \rangle + \langle Cov(M_0, M_2) \rangle \tag{12}$$

where the covariance term represents spatial correlations that enable equilibrium at $\langle M_C \rangle \neq 2$.

**Figure 1** demonstrates global synchronization to the frozen state. Despite different evolutionary paths dictated by local shear histories, trajectories converge to the same global equilibrium $M_C \approx 4.5$, providing crucial evidence of memory effects.

### *3.2 Asymptotic behavior*

The asymptotic behavior reveals the frozen state mechanism (**Figure 2**). Convective transport diminishes to machine precision ($\langle T_{conv} \rangle \sim 10^{-26}$), while diffusive transport becomes negligible ($\langle T_{diff} \rangle \sim 10^{-15}$). Coagulation continues at an extremely slow rate ($\langle T_{coag} \rangle \sim 10^{-9}$), consistent with flow decay factor $f(t)$ at late times ($t = 100$).

The system reaches a stalemate: flow becomes too weak to alter particle distribution, and coagulation rate (slaved to flow) is insufficient to drive toward homogeneous equilibrium. The asymptotic values confirm the frozen state:

$$\lim_{t \to \infty} \langle M_0 \rangle = 0.0362, \ \lim_{t \to \infty} \langle M_1 \rangle = 1, \ \lim_{t \to \infty} \langle M_2 \rangle = 124.3079, \ \lim_{t \to \infty} \langle M_C \rangle = 4.5026.$$

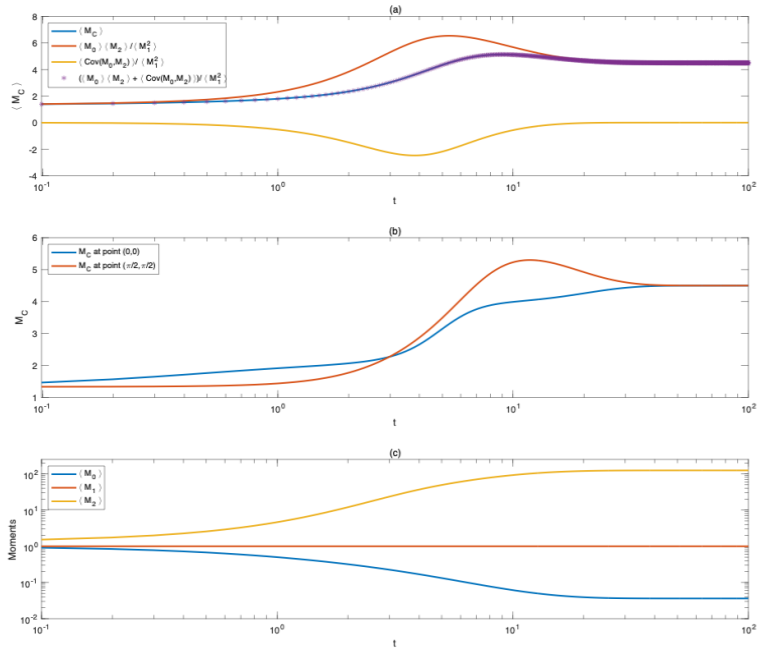

*Figure 1.* Approach to the frozen state under $Re = 10$, $Sc = 1$, $Da = 1$: (a) Composition of steady-state value of $\langle M_C \rangle \approx 4.5$ showing contributions from normalized second moment (red) and covariance term (yellow); (b) Evolution of $M_C$ at different locations demonstrating global synchronization; (c) Temporal evolution of moment fields confirming mass conservation.

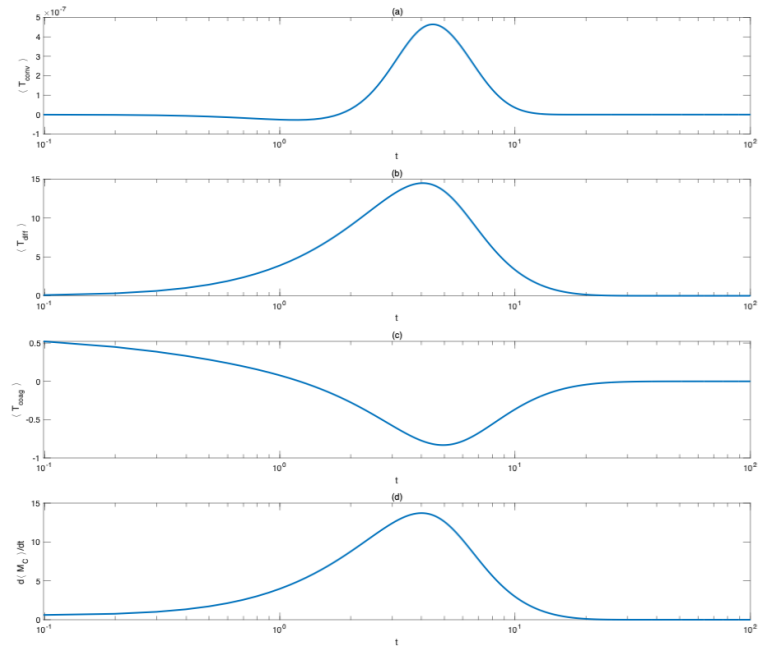

*Figure 2.* Asymptotic decay of transport contributions: (a) convective, (b) diffusive, (c) coagulative terms, and (d) evolution of $d\langle M_C \rangle/dt$ showing approach to steady state.

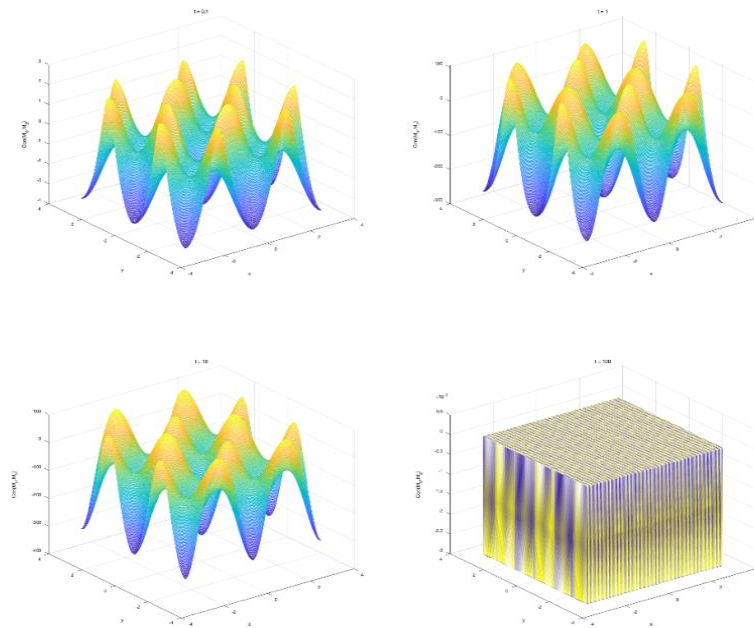

*Figure 3. Evolution of covariance term $Cov(M_0, M_2)$ at (a) $t = 0.1$; (b) $t = 1$;(c) $t = 10$ (d) $t = 100$ showing persistence of spatial correlations.*

### *3.3 Spatial structure and memory effects*

Analysis reveals that while individual moments become increasingly uniform, correlations between them persist (**Figure 3**). Early flow evolution establishes specific spatial patterns of particle sizes that become effectively frozen as flow decays, creating persistent memory of historical evolution.

At $t = 100$, the spatial distribution becomes nearly uniform with minor fluctuations of $Cov(M_0, M_2)$ ($\sim 10^{-3}$) approaching machine precision. This memory effect challenges the fundamental assumption that spatially averaged systems reduce to zero-dimensional models.

### *3.4. Implications and Applications*

**Theoretical implications:**

- **Breakdown of dimensional reduction:** Spatially extended systems cannot be reduced to zero-dimensional models for nonlinear processes like coagulation;
- **Historical dependence:** Final state depends on both current conditions and historical evolution;
- **Multiple equilibria:** Nonlinear systems with transport and reaction exhibit multiple stable equilibria.

**Practical applications:**

- **Atmospheric science:** Aerosol size distributions retain memory of earlier conditions;
- **Industrial processes:** Initial mixing conditions permanently influence final product characteristics;
- **Environmental engineering:** Coagulation-based treatment effectiveness depends on flow history.

### 4. Conclusions

We have discovered a novel frozen state phenomenon in shear-induced coagulation where $M_C$ stabilizes at 4.5, significantly deviating from the theoretically predicted of 2 for homogeneous systems. This state persists indefinitely, sustained by extremely slow coagulation ($\langle T_{coag} \rangle \sim 10^{-9}$) even as transport diminishes to machine precision. The underlying mechanism lies in persistent spatial correlations established during early flow evolution. As flow decays, these correlations become effectively frozen, creating permanent memory of historical development. This challenges conventional assumption about spatial averaging and reveals that historical evolution permanently influences final states.

The frozen state at $M_C \approx 4.5$ corresponds to an extremely broad particle size distribution (geometric standard deviation $\approx 3.4$, coefficient of variation $\approx 1.87$), representing significant departure from classical coagulation paradigm. Our discovery demands a paradigm shift in modeling particle-laden flows. Persistent memory effects necessitate new theoretical frameworks accounting for historical correlations, opening new avenues for understanding complex multiphase systems across diverse applications.

Future work should explore this phenomenon across different flow configurations, coagulation kernels, and boundary conditions, with focus on developing next-generation moment closure schemes that capture history-dependent, spatially correlated nature of turbulent coagulation.

**Acknowledgements**

This work was funded by the National Natural Science Foundation of China with grant number 11972169.

# Conflict of Interest

The authors declare no known competing financial interests or personal relationships that could have influenced the work reported in this paper.

Mingliang Xie (Corresponding Author)
Email: [mlxie@mail.hust.edu.cn](mailto:mlxie@mail.hust.edu.cn)